\documentclass[5p,10pt]{elsarticle}

\journal{Technical Report}

\usepackage[T1]{fontenc}
\usepackage{amsmath}
\usepackage{booktabs} %
\usepackage{multirow} %
\usepackage{paralist}

\usepackage{hyperref}
\usepackage{color}

\usepackage{graphicx}
\graphicspath{{./}{./figures/}}

\newcommand{\sstitle}[1]{\smallskip\noindent\textbf{#1.\/}}

\begin{document}

\setlength{\belowdisplayskip}{1pt}
\setlength{\belowdisplayshortskip}{1pt}
\setlength{\abovedisplayskip}{1pt}
\setlength{\abovedisplayshortskip}{1pt}

\begin{frontmatter}

\title{Technology Mapping with Large Language Models}

\author{Minh Hieu Nguyen}
\author{Hien Thu Pham}
\author{Hiep Minh Ha}
\author{Ngoc Quang Hung Le}
\author{Jun Jo}

\begin{abstract}
In today's fast-evolving business landscape, having insight into the technology stacks that organizations use is crucial for forging partnerships, uncovering market openings, and informing strategic choices. However, conventional technology mapping, which typically hinges on keyword searches, struggles with the sheer scale and variety of data available, often failing to capture nascent technologies. To overcome these hurdles, we present STARS (Semantic Technology and Retrieval System), a novel framework that harnesses Large Language Models (LLMs) and Sentence-BERT to pinpoint relevant technologies within unstructured content, build comprehensive company profiles, and rank each firm's technologies according to their operational importance. By integrating entity extraction with Chain-of-Thought prompting and employing semantic ranking, STARS provides a precise method for mapping corporate technology portfolios. Experimental results show that STARS markedly boosts retrieval accuracy, offering a versatile and high-performance solution for cross-industry technology mapping.
\end{abstract}

\end{frontmatter}

\section{Introduction}

In today's rapidly evolving business landscape, the emergence of new companies has been accelerating, driven by technological innovation and market competition. Identifying the technologies these companies are working on is crucial for creating business relationships, detecting market opportunities, and informing strategic decisions \cite{castells2001technology}. Technology mapping provides companies with the ability to visualize technological trends, understand competitors' strengths, and detect emerging areas of innovation \cite{castells2001technology}. Such insights enable businesses to stay competitive by aligning their research and development efforts with market demands and make informed decisions based on up-to-date technological intelligence~\cite{thang2022nature,duong2022efficient,nguyen2020factcatch,hung2017answer,nguyen2017argument}.

However, traditional methods for technology identification and mapping, such as keyword-based approaches, have limitations when dealing with large and heterogeneous datasets. They often lack the flexibility to adapt to emerging technologies and are domain-specific, which restricts their application across industries \cite{putthividhya2011bootstrapped}. As a result, there is a growing need for more sophisticated methods capable of processing vast amounts of unstructured data.

Recent advancements in Natural Language Processing (NLP) and Large Language Models (LLMs), such as GPT-3 and PaLM, have demonstrated their potential in overcoming these limitations. These models excel in zero-shot and few-shot learning, making them highly effective for extracting technology-related entities from unstructured data without requiring extensive labeled datasets \cite{brown2020language,wei2022emergent}. However, while LLMs can extract relevant entities, they cannot rank technologies effectively with specific companies.

To address this, we propose STARS (Semantic Technology and Retrieval System), a novel framework that combines LLM-based entity extraction with BERT-based semantic ranking. By leveraging BERT contextual embedding capabilities, STARS accurately matches technologies with companies, enabling a more precise and fine-grained understanding of their technological portfolios \cite{zhang2022semantic}. This approach efficiently maps the technology landscape across multiple industries, providing strategic insights for businesses and policymakers.
Our contributions are summarized as follows:
\begin{itemize}
    \item We introduce a method that integrates Chain-of-Thought prompting with LLMs to improve the extraction of relevant entities and technologies from unstructured data sources.
    \item We apply a semantic ranking technique using BERT to enhance the accuracy of matching technologies with companies, providing more precise, context-driven comparisons.
    \item We conduct a comprehensive evaluation demonstrating the scalability and retrieval precision of STARS in complex settings.
\end{itemize}

The rest of this paper is structured as follows: Section 2 provides an overview of the related works on technology extraction and ranking. In Section 3, we present the preliminaries, defining the problem and key concepts. Section 4 details our methodology, including the LLM-based entity extraction and BERT-based semantic ranking. Section 5 describes the experimental design, datasets, evaluation metrics, and presents the results of our experiments. Finally, Section 6 concludes the paper by summarizing the key findings and outlining potential directions for future research.

\section{Related Works}

The rapid expansion of big data has spurred researchers and practitioners to create various automated information retrieval methods, employing diverse yet often complementary approaches. These methods have been utilized extensively across areas such as digital libraries \cite{hersh2014biomedical}, search engines \cite{aggarwal2018machine}, media search \cite{rao2019multi} and recommender systems and information filtering \cite{Heggo2021}.

Many researches have studied how to extract and map technological trends. Previous research has also applied Named Entity Recognition (NER) and rule-based systems for extracting product attributes and values from listing titles \cite{putthividhya2011bootstrapped}. Aharonson and Schilling (2016) devised a technique to calculate the distance between patents and map out organizations' technological footprints \cite{aharonson2016mapping}. Likewise, Hossari et al. (2019) introduced an automated system for detecting emerging technologies within text-based documents \cite{hossari2019terminology}. Despite their usefulness, these approaches tend to be constrained to specific domains like patents and scientific articles and often struggle with handling large, heterogeneous datasets~\cite{zhao2021eires,huynh2021network,duong2022deep,nguyen2022model,nguyen2022detecting,trung2022learning,huynh2023efficient}. 

Recent advances have employed pre-trained language models (PLMs) such as BERT for entity extraction tasks, which have shown improved generalization compared to earlier methods~\cite{li2022mave}. Previous studies explored recommendation-based retrieval methods that map the relationship between companies and technologies, achieving notable success using models like DistilBERT~\cite{duong2023recommendation}. This approach focuses on the contextual similarity between entities, allowing for effective technology classification and retrieval in data-scarce environments. However, these approaches require large amounts of task-specific training data and struggle with generalizing to unseen attributes or technologies~\cite{nguyen2023poisoning,nguyen2023example,nguyen2014reconciling,nguyen2015smart,thang2015evaluation,nguyen2015tag,hung2019handling}.

In contrast, Large Language Models (LLMs) like GPT-3 and PaLM, pre-trained on vast amounts of text, have demonstrated the ability to overcome these limitations, excelling in zero-shot and few-shot learning tasks. Studies show that LLMs can achieve performance comparable to fine-tuned PLMs like BERT in extracting entity-technology pairs, even when provided with minimal examples \cite{brinkmann2023product}. LLMs have revolutionized various fields of natural language processing (NLP), particularly in tasks like entity extraction and information retrieval. Their ability to process and generate contextually accurate results through advanced techniques like few-shot and zero-shot learning has gained significant attention \cite{brown2020language,wei2022emergent}. The advent of techniques like Chain-of-Thought prompting enables LLMs to simulate a human-like reasoning process, improving the precision of extracted entities from unstructured documents \cite{wei2022emergent,kojima2022large}.

While LLMs can effectively extract relevant technologies, their ability to rank these technologies in relation to companies is limited. To address this, semantic ranking techniques, particularly BERT, have been used to compute the similarity between company descriptions and technology definitions. Semantic ranking models have been widely applied in recommendation systems and information retrieval, offering a nuanced understanding of relationships between entities by embedding them in a shared vector space \cite{zhang2022semantic}. Zhang et al. \cite{zhang2022oa} used a BERT-based approach to rank attribute-value pairs extracted from product titles, further highlighting the effectiveness of semantic ranking~\cite{yang2024pdc,sakong2024higher,huynh2024fast,huynh2025certified,nguyen2023isomorphic,nguyen2024portable}.

\section{Preliminaries}

In this section, we formalize the problem of technology mapping and provide mathematical notations and definitions that are essential for our methodology.

\subsection{Problem Definition}

Let $\mathcal{C} = \{c_1, c_2, \dots, c_m\}$ be a set of companies, and let $\mathcal{T} = \{t_1, t_2, \dots, t_n\}$ be a set of technologies. Our goal is to extract relevant technologies from unstructured data for each company $c_i \in \mathcal{C}$, and rank them based on their relevance. Formally, we aim to map each company $c_i$ to a subset of technologies $\hat{\mathcal{T}}_i \subseteq \mathcal{T}$, where $\hat{\mathcal{T}}_i$ represents the most relevant technologies for $c_i$.

The technology mapping problem can be broken down into two main tasks:
\begin{compactitem}
    \item \textbf{Entity extraction:} For each company $c_i$, extract relevant technologies from unstructured text data. This involves identifying potential technologies $\tilde{\mathcal{T}}_i$ from a corpus of documents $D_i$ related to $c_i$.
    \item \textbf{Semantic ranking:} Once $\tilde{\mathcal{T}}_i$ is extracted, we rank the technologies in $\tilde{\mathcal{T}}_i$ such that the top-$k$ technologies are the most relevant to $c_i$.
\end{compactitem}

\subsection{Mathematical Formulation}

Let $x_i \in D_i$ be a document related to company $c_i$. The goal of the entity extraction task is to identify a set of potential technologies $\tilde{\mathcal{T}}_i$ from $x_i$. Formally:
\begin{equation}
\tilde{\mathcal{T}}_i = \arg\max_{\mathcal{T}} P(\mathcal{T} | x_i),
\end{equation}
where $P(\mathcal{T} | x_i)$ represents the probability of a technology $\mathcal{T}$ being relevant to the document $x_i$. The Chain-of-Thought prompting technique we use improves the reasoning process for the Large Language Model (LLM), allowing it to better infer technologies even when they are not explicitly mentioned.

Once the set $\tilde{\mathcal{T}}_i$ is extracted, the next step is to rank these technologies based on their relevance to company $c_i$. Let $\mathbf{e}_{c_i}$ be the embedding of company $c_i$, and $\mathbf{e}_{t_j}$ be the embedding of technology $t_j$. We use a BERT-based semantic ranking approach to compute the similarity score $S(c_i, t_j)$ between company $c_i$ and each technology $t_j \in \tilde{\mathcal{T}}_i$:
\begin{equation}
S(c_i, t_j) = \frac{\mathbf{e}_{c_i} \cdot \mathbf{e}_{t_j}}{\|\mathbf{e}_{c_i}\| \|\mathbf{e}_{t_j}\|}.
\end{equation}

The final ranked set of technologies $\hat{\mathcal{T}}_i$ is determined by selecting the top-$k$ technologies with the highest similarity scores.

\subsection{Challenges in Technology Mapping}

Technology mapping involves several key challenges:
\begin{itemize}
    \item \textbf{Data diversity:} The unstructured documents related to companies vary in format (e.g., web pages, patent filings, job postings) and content. This requires a flexible and robust extraction method capable of generalizing across different types of data.
    \item \textbf{Emerging technologies:} Companies often work with emerging technologies that may not be well-documented or captured in predefined lists. This necessitates a model that can infer technologies from context, rather than relying solely on explicit mentions.
    \item \textbf{Contextual relevance:} Technologies may have varying levels of relevance to a company  operations. A key challenge is ranking technologies not only based on their presence in the company  documents but also on their strategic importance to the company.
\end{itemize}

Our approach addresses these challenges by leveraging Chain-of-Thought prompting for improved inference during entity extraction and using semantic ranking to ensure that technologies are ranked based on their contextual relevance to each company.

\section{Methodology}

\begin{figure*}[!h]
    \centering
    \includegraphics[width=1\textwidth]{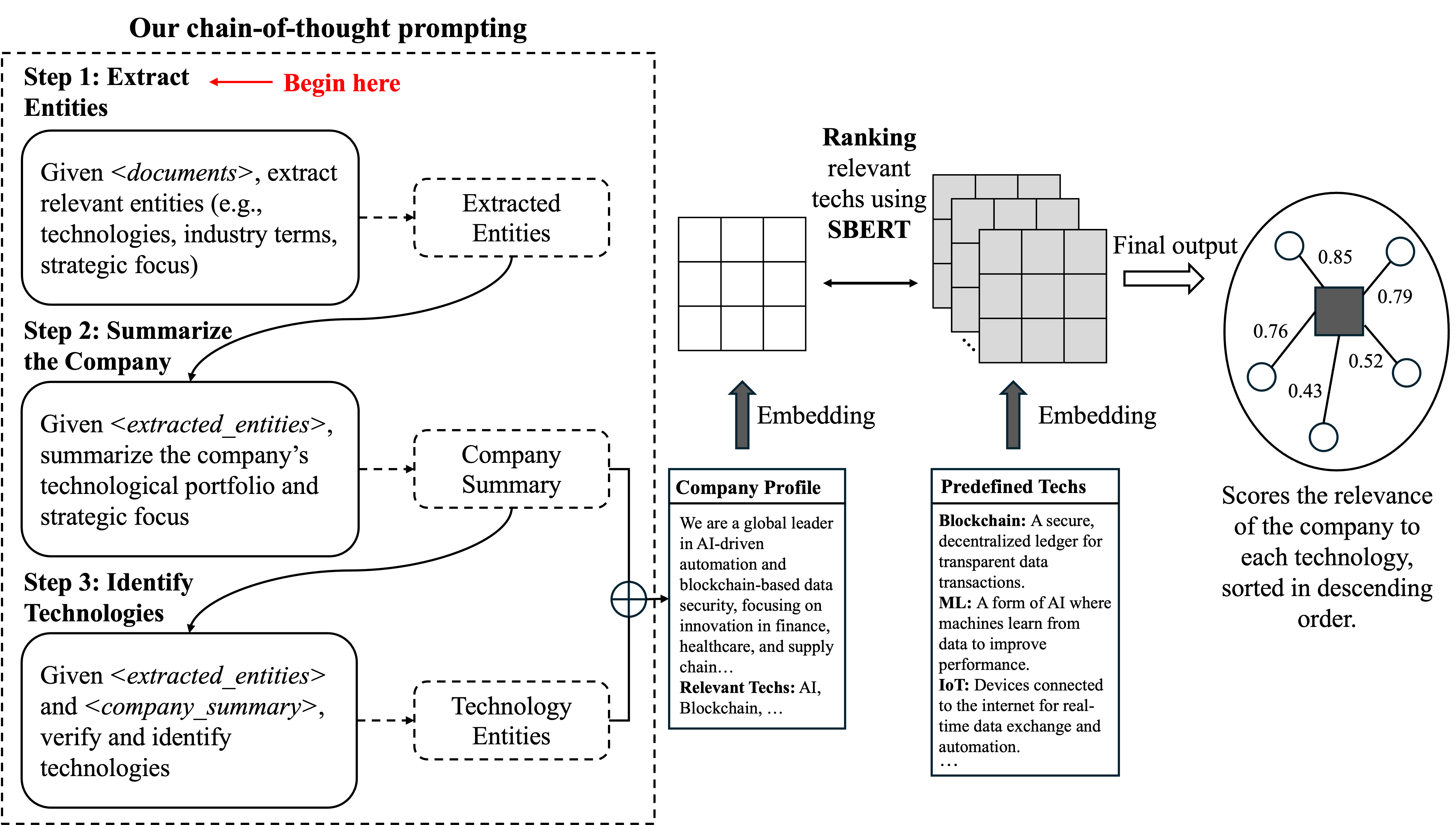}
    \caption{Overview of the Framework for Technology Landscape Mapping: STARS.}
    \label{fig:framework}
\end{figure*}

In this section, we present our methodology for mapping the technology landscape, which is visually summarized in Figure \ref{fig:framework}. Our approach consists of three key components: (1) Entity Extraction using LLMs with Chain-of-Thought Prompting, (2) Company Summarization, and (3) Technology-Company Retrieval via Semantic Ranking using Sentence-BERT. This methodology leverages the power of LLMs to extract relevant entities such as technologies and companies from unstructured text, generate comprehensive company profiles, and accurately rank technologies based on their relevance to a company  profile. By utilizing the LLM's capabilities to identify both existing and emerging technologies that may not be captured in predefined datasets, we ensure that the process is both scalable and adaptable to various industries and domains~\cite{nguyen2019maximal,nguyen2021judo,nguyen2020factcatch,nguyen2022survey,nguyen2024manipulating}.

\subsection{Entity Extraction with Chain-of-Thought Prompting}

The first phase involves extracting relevant entities from a corpus of unstructured documents. Extracting entities such as technologies, companies, and innovations is a critical step in mapping the technological landscape. As highlighted by Duong et al. \cite{duong2023recommendation}, identifying entities from raw, unstructured data provides a structured representation, enabling us to map the relationships between companies and the technologies they leverage. Without this extraction process, identifying implicit or emerging technologies linked to companies becomes nearly impossible due to the vast amount of data and the complexity of implicit connections between companies and their innovations.

Entity extraction allows us to identify specific technologies linked to a company, which might not be explicitly mentioned but inferred from broader contexts. By identifying and classifying these entities, we can systematically associate them with companies. For instance, when we extract entities like \emph{blockchain} or \emph{machine learning} from documents, we can infer these technologies are relevant to the company even if not directly stated. This process aligns with Duong et al. approach of constructing a technology-company interaction matrix, where both explicit and inferred entities are used to build a comprehensive map of company activities and their technological portfolio \cite{duong2023recommendation}.

Moreover, extracting entities enables us to map relationships across a broad set of companies and technologies, facilitating technology discovery and company profiling. This step is essential, as companies often engage with new or evolving technologies that may not yet be captured in predefined databases but can be inferred from related entities. By systematically extracting and contextualizing entities, we create a foundation for deeper analysis, such as semantic ranking and retrieval of company-technology relationships~\cite{nguyen2015result,nguyen2017retaining,tam2019anomaly,nguyen2020monitoring,nguyen2019user,nguyen2020entity,nguyen2021structural}.

\sstitle{Chain-of-Thought Prompting}
Chain-of-Thought (CoT) prompting guides the LLM through a series of reasoning steps, simulating a logical thought process similar to that of a human. The CoT prompting is defined by a sequence of prompts, $\mathbf{P} = \{p_1, p_2, \dots, p_n\}$, where each prompt $p_i$ is conditioned on the output of the previous one, thereby creating a chain of logical inferences.

Mathematically, the entity extraction process can be represented as:

\begin{equation}
\hat{y} = \arg\max_{y \in \mathcal{Y}} P(y | x, \mathbf{P}),
\end{equation}
where $\hat{y}$ is the predicted entity, $x$ is the input document, and $\mathcal{Y}$ is the set of potential entities. The probability $P(y | x, \mathbf{P})$ is computed by the LLM, considering the chain of thought prompts $\mathbf{P}$ that progressively refine the context for evaluating $y$.

The CoT approach allows the LLM to consider a broader context, extracting entities that may not be explicitly mentioned but are inferred from the surrounding text. For instance, in a document discussing a company  innovations, the LLM might infer that terms like "deep learning" and "AI" are technologies relevant to the company, even if these terms are not explicitly listed.

\begin{figure*}[!h]
    \centering
    \includegraphics[width=1\textwidth]{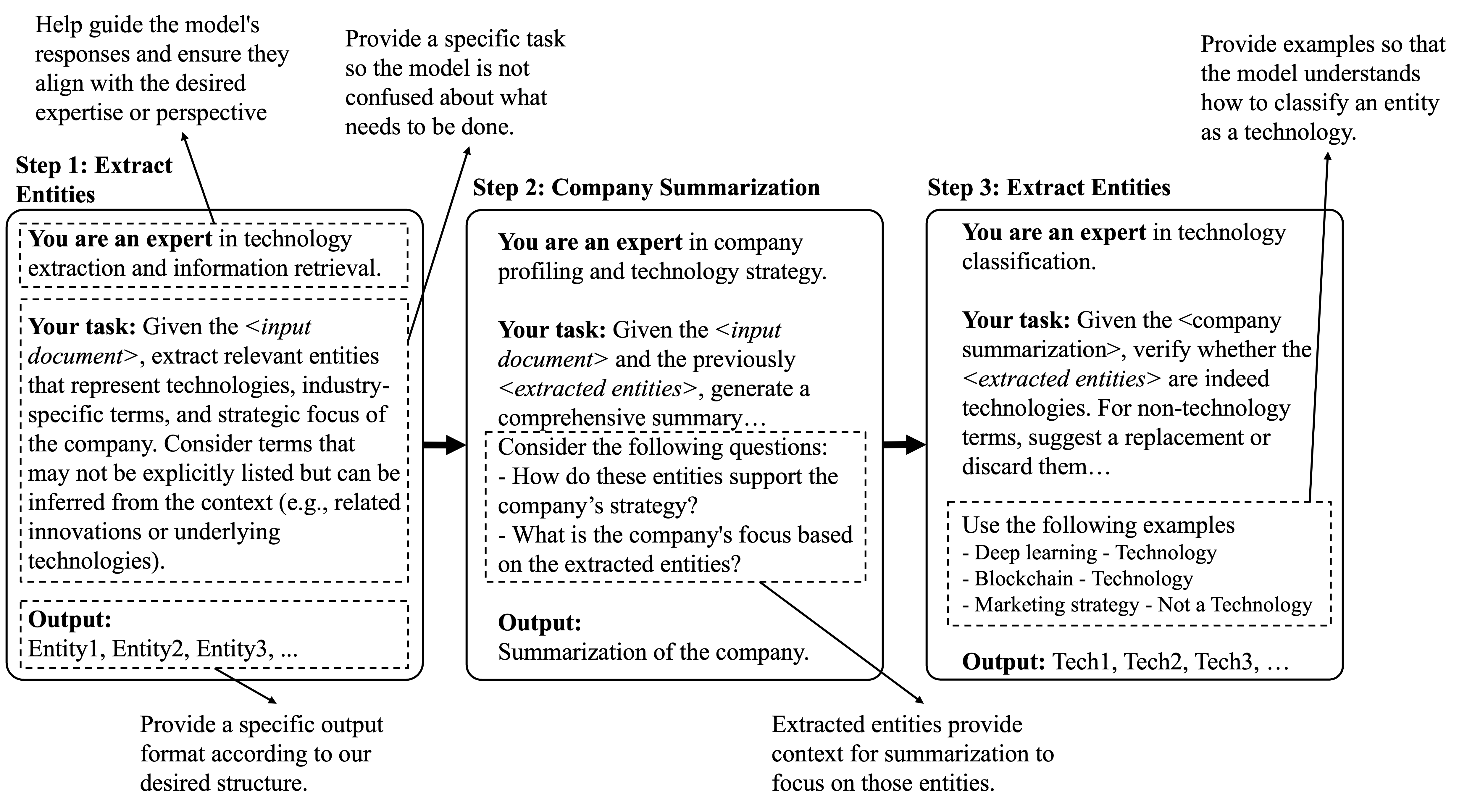}
    \caption{Overview of our Chain-of-Thought Prompts for Technology Extraction.}
    \label{fig:prompt}
\end{figure*}

Figure \ref{fig:prompt} describes the Chain-of-Thought prompting process, which guides the LLM to extract relevant entities, summarize company profiles, and classify technologies effectively. This prompt structure provides explicit guidance and examples, ensuring that the model produces contextually accurate results aligned with the company's technological portfolio.
The CoT prompt is designed with the following steps:

\begin{compactitem}
\item \textbf{Step 1: Extract Entities.} The LLM identifies and extracts all relevant entities from the input document, focusing on potential technologies. After extracting relevant entities, the next step is to summarize the company  technological portfolio. The extracted entities from this step serve as the input for this summarization. By analyzing the entities related to a company (such as key technologies, industry terms, and strategic focus), the LLM is able to generate a contextualized summary that reflects the company's operations and technological strengths.

\item \textbf{Step 2: Summarize the Company.} The LLM generates a comprehensive summary of the company, highlighting its technological portfolio and strategic focus. This summary incorporates insights into the company's technological strengths, areas of innovation, and potential opportunities. 
The purpose of company summarization is to provide a coherent and concise overview of a company  technological focus, which is crucial for the subsequent ranking of technologies. Summarization adds context to the entities extracted in Step 1, allowing us to better understand how the extracted technologies relate to the company  broader goals and operations. A summary helps create a narrative that connects the individual entities, providing the necessary background for accurate technology ranking. For instance, after extracting technologies such as "AI" and "blockchain" for a given company, the summarization step helps identify how these technologies are integrated into the company  operations. This might involve generating insights about the company's product lines, R\&D efforts, or strategic goals that are linked to these technologies.

\item \textbf{Step 3: Identify Technologies.} After extracting entities in Step 1 and summarizing the company information in Step 2, the LLM verifies whether the entities are indeed technologies. This is done by comparing them against known technology definitions or by leveraging its internal knowledge of technology concepts.

\end{compactitem}

We also employ few-shot example prompting, where the LLM is provided with a small set of examples that demonstrate what constitutes a technology. This method not only relies on predefined labels but also enhances the LLM's ability to generalize from a few examples. For instance:

\begin{verbatim}
Example 1: Deep learning - Technology
Example 2: Blockchain - Technology
Example 3: Marketing strategy - Not a Technology
\end{verbatim}
To enhance this process, we utilize a labeled dataset of technologies derived from a prior study \cite{duong2023recommendation}.
The dataset was built by analyzing Wikipedia  Main Topic Classifications (MTCs) \footnote{\url{https://en.wikipedia.org/wiki/Category:Main_topic_classifications}}, focusing on Technology, Science, and Engineering. A top-down approach cleaned irrelevant entries like admin pages and companies, linking categories to MTCs based on the shortest path. Ultimately, 1,356 categories were manually labeled as technologies, providing a strong foundation for technology classification.

By using example prompts and the labeled dataset from this prior study, the LLM generalizes and classifies new entities as either technologies or non-technologies, making this approach highly efficient while reducing the need for extensive training data.

After generating the extracted technology identities and company summary, we can identify which technologies are relevant to a company based on a predefined list. This ranking process is explained in detail in the next section, \textit{Technology-Company Retrieval via Semantic Ranking}.

\subsection{Technology-Company Retrieval via Semantic Ranking}

The second phase involves determining which technologies are most relevant to a given company. After extracting the company summary and relevant entities (technologies), the challenge lies in identifying the technologies that are most pertinent to the company's operations. While the LLM is capable of suggesting relevant technologies based on the extracted information, it does not inherently rank these technologies. Without a ranking mechanism, it can be difficult to evaluate the relative importance or relevance of the technologies for the company.

\sstitle{Sentence-BERT for Semantic Ranking}
Sentence-BERT (SBERT) is a modification of the BERT network, specifically designed for producing semantically meaningful sentence embeddings. These embeddings can be used for tasks like clustering, semantic search, and ranking. Unlike the original BERT, which is not optimized for semantic similarity, SBERT creates embeddings that allow for faster and more accurate comparisons between textual entities. For example, finding the most similar sentence pair in a collection of 10,000 sentences would take BERT around 65 hours, while SBERT reduces this process to approximately 5 seconds \cite{reimers2019sentencebert}.

To perform semantic ranking, we first embed the technologies using a pre-trained Sentence-BERT model. Let $\mathbf{e}_{t_j}^{SBERT}$ denote the SBERT embedding of technology $t_j$. These embeddings are computed based on the technology name and its corresponding definition. Specifically, for each technology $t_j$, the embedding is calculated as:
\begin{equation}
\mathbf{e}_{t_j}^{SBERT} = g(\text{tech\_name}(t_j), \text{definition}(t_j)),
\end{equation}
where $g(\cdot)$ is the Sentence-BERT model that generates an embedding by jointly encoding the technology name $\text{tech\_name}(t_j)$ and its definition $\text{definition}(t_j)$. Definitions are obtained either from Wikipedia or generated by the LLM, leveraging its vast training corpus to produce contextually accurate and rich definitions for technologies not well-documented in external sources.

Next, we generate a company profile embedding $\mathbf{e}_{c_i}^{SBERT}$ by combining the company's summary (extracted by the LLM) with the embeddings of the identified technologies:

\begin{equation}
\mathbf{e}_{c_i}^{SBERT} = f(\mathbf{e}_{c_i}, \mathbf{e}_{t_1}^{SBERT}, \dots, \mathbf{e}_{t_k}^{SBERT}),
\end{equation}
where $f(\cdot)$ is a function that aggregates the embeddings, summarizing the company  technological profile.

We then compare the company  profile embedding $\mathbf{e}_{c_i}^{SBERT}$ with the predefined technology embeddings $\mathbf{e}_{t_j}$ using cosine similarity suggested by the LLM:

\begin{equation}
S_{rank}(c_i, t_j) = \frac{\mathbf{e}_{c_i}^{SBERT} \cdot \mathbf{e}_{t_j}^{SBERT}}{\|\mathbf{e}_{c_i}^{SBERT}\| \|\mathbf{e}_{t_j}^{SBERT}\|}.
\end{equation}
This similarity score, $S_{rank}(c_i, t_j)$, allows us to rank the technologies. The final ranked list of technologies for each company is determined by selecting the top-$k$ technologies with the highest semantic similarity scores. Using Sentence-BERT for this task significantly enhances semantic ranking precision and reduces computation time compared to traditional BERT-based ranking approaches \cite{reimers2019sentencebert}.

In the end, we obtain a ranked list of technologies that are most relevant to the company, based on the semantic similarity between the company's profile and predefined technology embeddings. By leveraging Sentence-BERT for embedding and ranking, we enhance the precision of the technology retrieval process, ensuring that the final ranked technologies are accurate.
\section{Experiments}

In this section, we detail our experimental setup, datasets, evaluation metrics, and results for both the end-to-end and semantic ranking evaluations. We employ Precision at K (P@k) as our primary metric to assess retrieval accuracy. The experiments are designed to evaluate the impact of different prompting methods and the effectiveness of SBERT-based ranking.

\subsection{Datasets}

The datasets used for our experiments were compiled from multiple publicly available sources, including company websites, patent databases, and job postings. To create a focused dataset, we started by obtaining a comprehensive list of industries from Crunchbase \footnote{\url{https://support.crunchbase.com/hc/en-us/articles/27690673553555-Glossary-of-Industries}}. From this list, we manually filtered and selected only the industries categorized as technology-related, resulting in a final set of 176 unique technologies.

For each of these 176 technologies, we crawled data from 50 companies appearing on the first page of Crunchbase that were categorized under each technology. This resulted in a dataset of 6,597 companies, representing a diverse array of technological industries. This diverse dataset allows us to evaluate the generalizability of our approach across different sectors and technologies.

\subsection{Metric: Precision at K (P@k)}

We use Precision at K (P@k) as the primary metric to evaluate our retrieval system. P@k measures the proportion of correct predictions within the top-$k$ retrieved results.
Let $R(c_i)$ be the set of relevant technologies for company $c_i$ and $\hat{R}_k(c_i)$ be the set of top-$k$ technologies retrieved by the system for company $c_i$. Then Precision at K is defined as:
\begin{equation}
P@k = \frac{1}{|\mathcal{C}|} \sum_{i=1}^{|\mathcal{C}|} \frac{|R(c_i) \cap \hat{R}_k(c_i)|}{|\hat{R}_k(c_i)|},
\end{equation}
where $|\mathcal{C}|$ is the number of companies, $R(c_i)$ represents the set of relevant technologies for company $c_i$, and $\hat{R}_k(c_i)$ represents the top-$k$ retrieved technologies for company $c_i$. The same formula is applied for the reverse task of technology-to-company matching.

After analyzing the distribution of technologies across these companies, we found that a company can have a maximum of 11 associated technologies, though only a few companies have this many. Based on this statistical analysis, we chose to evaluate our system using $k=3, 5, 7, 10$, which represents different levels of precision for the top retrieved results.

\subsection{Evaluation}

\begin{figure}[!h]
\centering
\begin{minipage}{0.49\linewidth}
    \centering
    \includegraphics[width=\linewidth]{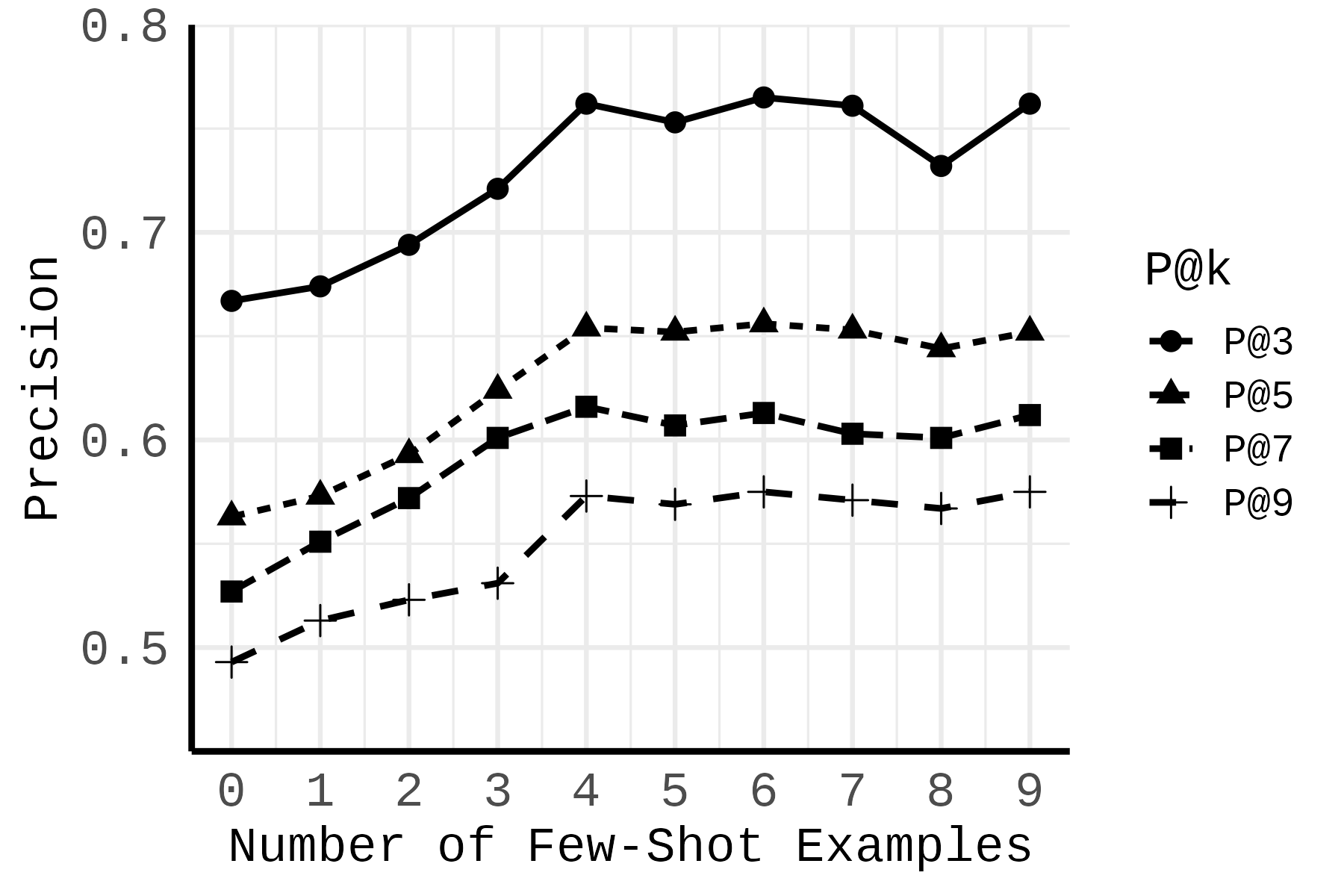}
    \caption{Effect of few-shot examples on P@k for company-to-technology retrieval.}
    \label{fig:few_shot}
\end{minipage}
\hfill
\begin{minipage}{0.49\linewidth}
    \centering
    \includegraphics[width=\linewidth]{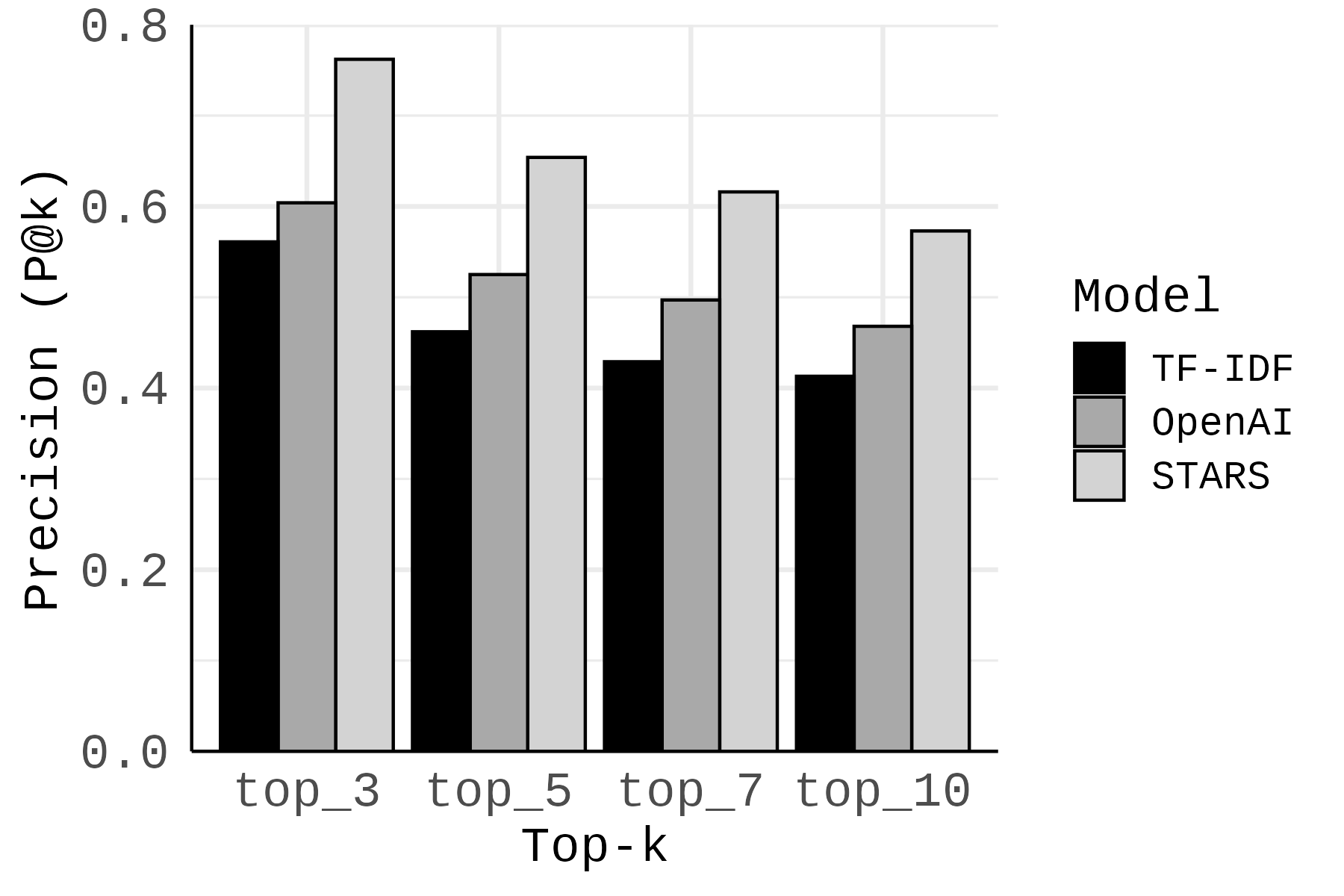}
    \vspace{-2em}
    \caption{Comparison of ranking methods for semantic matching. SBERT outperforms across all $k$ values.}
    \label{fig:ranking_score}
\end{minipage}
\end{figure}

\sstitle{Prompting Evaluation}
The prompting evaluation tests two settings:
\begin{itemize}
    \item \emph{Company-to-Technology Retrieval (Com-Tech):} evaluates the system  ability to retrieve relevant technologies for a given company.
    \item \emph{Technology-to-Company Retrieval (Tech-Com):} assesses how well the system retrieves companies that are associated with specific technologies.
\end{itemize}
The goal is to assess how well our LLM-based approach, combined with various prompting methods, performs in both retrieving technologies for a given company and matching companies with technologies. The results for different prompting methods are shown in Table \ref{tab:prompt_retrieval}. For all experiments, we use SBERT for ranking, as the results from Figure \ref{fig:ranking_score} indicate that it provides superior performance compared to other ranking methods.

\begin{table}[!h]
\centering
\footnotesize
\caption{Effects of prompting techniques.}
\label{tab:prompt_retrieval}
\scalebox{0.75}{
\begin{tabular}{lcccccccc}
\toprule
& \multicolumn{4}{c}{Company-Technology retrieval} & \multicolumn{4}{c}{Technology-Company retrieval} \\
\cmidrule(lr){2-5}\cmidrule(lr){6-9}
Model & top-3 & top-5 & top-7 & top-10 & top-3 & top-5 & top-7 & top-10 \\
\midrule
Single Prompt & 0.583 & 0.554 & 0.507 & 0.469 & 0.582 & 0.515 & 0.486 & 0.423 \\
CoT prompting & 0.667 & 0.563 & 0.527 & 0.493 & 0.628 & 0.556 & 0.503 & 0.457 \\
\textbf{STARS} & \textbf{0.762} & \textbf{0.654} & \textbf{0.616} & \textbf{0.573} & \textbf{0.725} & \textbf{0.634} & \textbf{0.588} & \textbf{0.549} \\
\bottomrule
\end{tabular}
}
\end{table}

As shown in Table \ref{tab:prompt_retrieval}, STARS consistently outperforms both Single Prompt and Chain-of-Thought (CoT) prompting techniques across all retrieval tasks. In both company-to-technology and technology-to-company settings, STARS achieves the best results. For instance, in company-to-technology retrieval, STARS reached a precision of 0.762 at the top-3 level, representing a 14.2\% improvement over CoT and 30.7\% over Single Prompt. Similar gains are observed in technology-to-company retrieval, where STARS shows a 15.5\% improvement over CoT and 24.6\% over Single Prompt at the top-3 level. Even at higher top-k levels, STARS maintains its advantage across both settings, with up to a 19.7\% increase in precision compared to CoT and a 29.2\% increase over Single Prompt. These results underscore STARS' ability to deliver more accurate and contextually relevant retrievals in various scenarios.

\sstitle{Few-Shot Evaluation}
We also assess the impact of few-shot learning by varying the number of few-shot examples used during prompting. Figure \ref{fig:few_shot} shows the effect of increasing the number of few-shot examples on system performance.

In the company-to-technology retrieval setting, as shown in Figure \ref{fig:few_shot}, increasing the number of few-shot examples results in a steady improvement in precision across all P@k levels. Starting from zero examples, the precision for P@3 is 0.667 and rises to 0.762 at five examples, reflecting a 14.3\% increase. This trend is consistent for P@5, P@7, and P@9, where precision improves notably up to five examples. For instance, P@5 increases from 0.563 to 0.654, and P@7 increases from 0.527 to 0.616. However, beyond five examples, the precision stabilizes with marginal fluctuations. For example, P@3 peaks at 0.765 with seven examples before leveling off to 0.762 at nine examples. These results suggest that the optimal balance between model improvement and diminishing returns is observed at five examples, highlighting its importance for effective retrieval without overfitting.

\sstitle{Semantic Ranking Evaluation}
For the semantic ranking evaluation, we compare SBERT with other simpler methods, such as TF-IDF (Term Frequency Inverse Document Frequency) embeddings and scores generated by ChatGPT. The SBERT consistently achieves the highest precision, as it captures deeper contextual relationships between companies and technologies.

As shown in Figure \ref{fig:ranking_score}, STARS consistently achieves the highest precision, outperforming both OpenAI and TF-IDF across all values of $k$. For instance, at the top-3 level, STARS reaches a precision of 0.762, compared to 0.604 for OpenAI and 0.561 for TF-IDF. This trend continues at the top-5, top-7, and top-10 levels, with STARS maintaining its advantage with precisions of 0.654, 0.616, and 0.573, respectively. These results highlight the ability of STARS to capture deeper contextual relationships between companies and technologies, while OpenAI and TF-IDF, though effective, fall behind due to their more generalized or simpler approaches.

\section{Conclusion}
This paper proposed STARS, a framework that combines LLM-based entity extraction with BERT-based semantic ranking for mapping technologies to companies. By using Chain-of-Thought prompting and Sentence-BERT, STARS enhances precision in technology retrieval from unstructured data, offering a scalable solution across industries. Experiments showed that STARS outperforms traditional methods in both retrieval tasks, with notable improvements in precision using few-shot learning. Future directions include graph learning~\cite{lee2024sfgcn}, privacy consideration~\cite{zhang2024does,liu2024matrix}, and recommender systems~\cite{nguyen2024manipulating,nguyen2025privacy,pham2024dual,nguyen2024multi,nguyen2024handling}.

%\bibliographystyle{elsarticle-num-names}
%\bibliography{../mybibliography,../ref_h}

\end{document}